\documentclass[prl,aps,twocolumn,groupedaddress,floats,showpacs]{revtex4}
\usepackage{graphicx}
\usepackage{dcolumn}
\usepackage{bm}
\usepackage{color}
\def\he4{$^4$He}
\def\hel3{$^3$He}
\def\Am3{\AA$^{-3}$}
\def\beq{\begin{equation}}
\def\eeq{\end{equation}}
\begin{document}
\author{A. B. Kuklov}
\affiliation{Department of Physics \& Astronomy of the College of Staten Island, and the Graduate Center,
CUNY}


\title{Plasticity induced superclimb in solid Helium-4: Direct and inverse effects.}

\date{\today}
\begin{abstract}
During the last decade the experimental evidence is building that the mass supertransport through solid \he4 as well as the anomalously large matter accumulation in the bulk -- the giant isochoric compressibility (aka the syringe effect) -- are both supported by a network of dislocations with superfluid core. However, a structure of this network  as well as its relation to the basal (non-superfluid) dislocations which are responsible for plasticity remain unclear.  Here it is shown that superclimbing and basal edge dislocations can form bound pairs. This implies that plastic deformation should produce the syringe effect and vice versa.  The experimental test is proposed.  While the strength of the effect depends on the average orientation of the paired dislocations, there is a feature unique to the superfluid dislocation scenario -- the supercurrents flow in the direction perpendicular to the plastic deformation.
\end{abstract}



\maketitle

Superflow through solid \he4 as well as the syringe effects  have been discovered in UMASS group \cite{Hallock}.
While the strength of the flow was extremely small (about few ng/s), the amount of matter accumulated inside the bulk has indicated that the solid exhibited the response on the applied chemical potential as large as that of a liquid.  The principal features of the effects have been confirmed by two other groups \cite{Beamish,Moses}. In the  experiment \cite{Beamish} the intrinsic flow from one part of solid \he4 to another has been found, with the rate increasing as temperature lowered. This behavior excluded any explanation of the syringe effect within classical plasticity. Temperature, pressure and bias dependencies of the superflow through solid have been studied in detail in Ref. \cite{Moses}. These turn out to be consistent with the original observations \cite{Hallock}. Furthermore, in Ref.\cite{Moses} an explanation in terms of possible macroscopic liquid channels (existing along the boundaries between a sample and walls and responsible for the superflow) has been excluded, and it was concluded that the superflow through solid \he4 occurs through a network of superfluid dislocations observed in {\it ab initio} simulations \cite{screw,sclimb}. It is important to note that the temperature dependence of the flow at $T>0.1-0.2$K is essentially insensitive to the orientation of the crystal \cite{Moses}. This indicates that the dislocation network is mainly uniform and isotropic -- that is, it consists of comparable numbers of segments of screw \cite{screw} and edge \cite{sclimb} types.    

That a network of dislocations with superfluid core represents a system with unique dynamical properties has been pointed out in Ref.\cite{shevchenko} long before the observations \cite{Hallock,Beamish,Moses}.  This model, however, does not take into account the superclimb -- that is, a climb of edge dislocations with superfluid core resulting in the syringe effect \cite{sclimb}. An unusual feature is that the syringe effect is essentially independent of the density of the superfluid dislocations as long as their network is uniform over the solid \cite{sclimb}.  
Observing such a feature would be a direct confirmation of the superfluid dislocation scenario. However, an imaging of dislocations in solid \he4 simultaneously with measuring the syringe effect does not appear to be possible. Here another experiment is proposed to serve as a "smoking gun" for the superfluid dislocation network scenario as a basis for the observations \cite{Hallock,Beamish,Moses}.      

The proposed experiment is based on measuring the syringe effect in response to the shear stress. At this point it is important to mention that the effect dubbed {\it supershear} has been proposed in Ref.\cite{ALKS}. It is analogous to the high temperature plasticity of granular media where the activated transport of vacancies along the grain boundaries (Coble plasticity \cite{Coble}) is replaced by superflow along the superfluid grain boundaries \cite{GB}. 
While representing one option for the interrelation between plasticity and superflow through solid, it cannot occur in a non-granular solid. Furthermore,  the boundary currents induced by shear are along the applied stress which will make this mechanism hard to distinguish  from the conservative glide of dislocations realizing the conventional plasticity \cite{Landau,Hirth}. In contrast to the supershear \cite{ALKS} which can be viewed as the longitudinal effect (with respect to the directions of strain and superflow), 
the one discussed below accounts for the transverse response on the applied shear -- that is, the superflow in the direction perpendicular to the applied shear.  Thus, this effect can be dubbed as {\it transverse supershear}.
\begin{figure}[!htb]
	\includegraphics[width=1.0 \columnwidth]{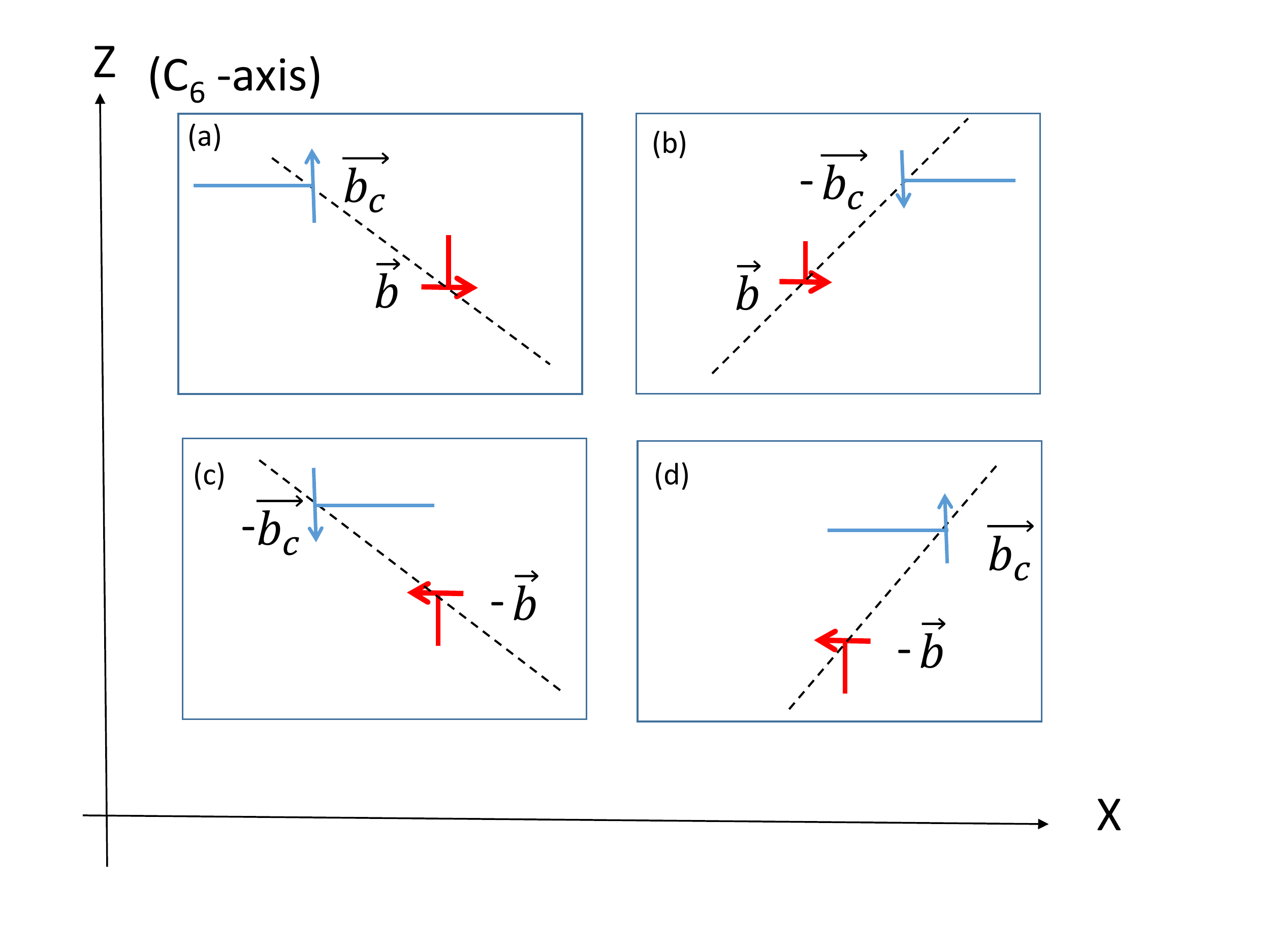}
	\vskip-8mm
	\caption{(Color online) Stable equilibrium positions (dashed lines) of basal gliding dislocation (red) bound to superclimbing one (blue). Each bound pair in (a),(b),(c),(d) cases is characterized by specific orientations of the Burgers vectors $\pm \vec{b}$ and $\pm \vec{b}_c$. The cores are aligned with the Y-axis (into the page) of the basal plane. Arrows indicate orientations of the Burgers vectors and the solid lines attached to them outline the half planes of extra atoms. Motion of the cores can only occur along the X-axis, while the superflow occurs along the Y-axis.  }
	\label{fig1}
\end{figure}
\begin{figure}[!htb]
	\includegraphics[width=1.0 \columnwidth]{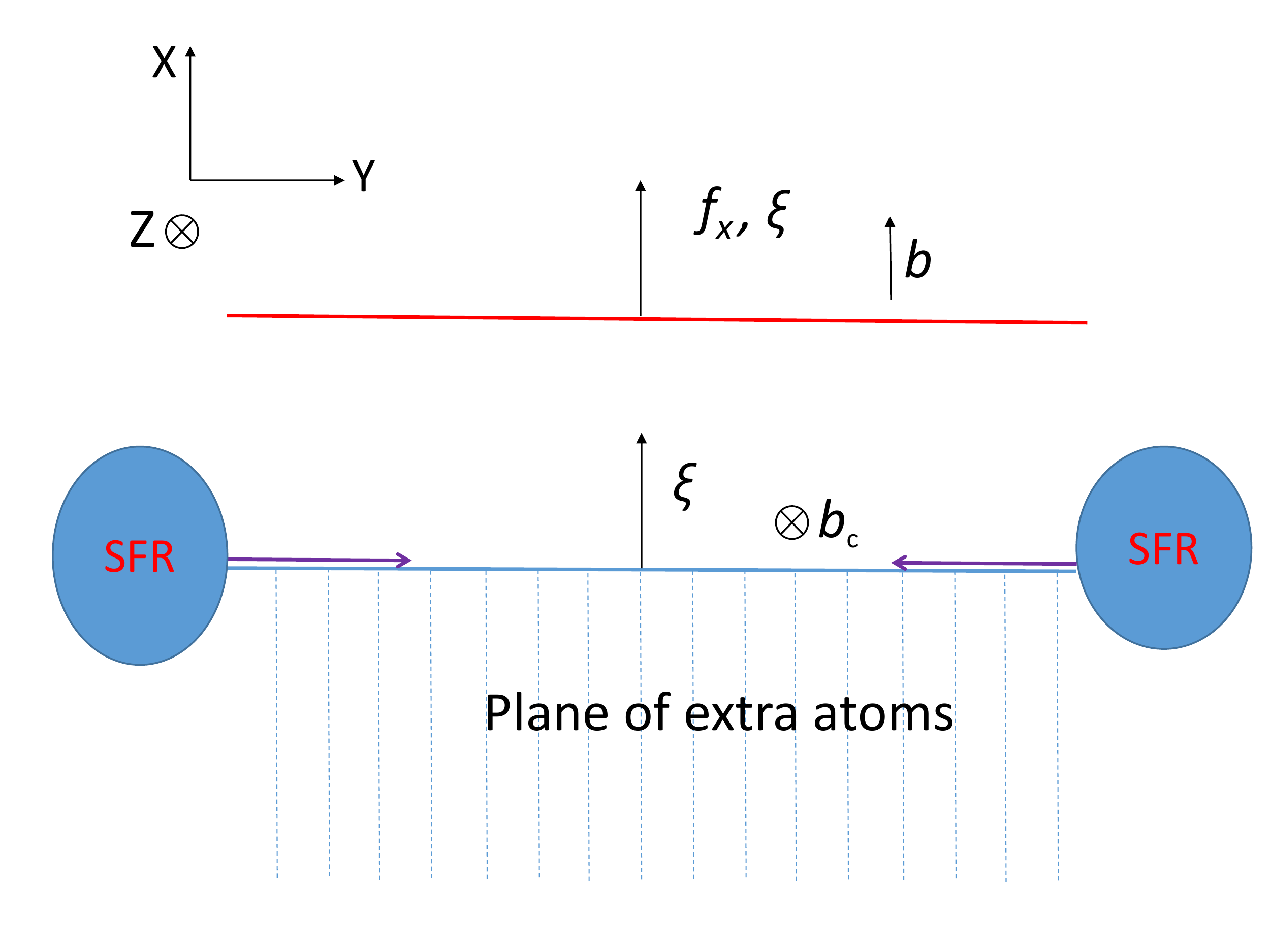}
	\vskip-4mm
	\caption{(Color online) Schematics of the transverse supershear effect. The view is along Z-axis. The upper and lower horizontal lines represent the basal (in red) and the superclimbing (in blue) dislocations, respectively. The directions of their Burgers vector, $b, b_c$, as well as the applied force $f_x$ and the resulting displacement $\xi$ are also shown. Two ellipses labeled as SFR represent reservoirs with superfluid, with the arrows along the core indicating supercurrents driven from the reservoirs by either the external force $f_x=f^{(ex)}\sim \sigma_{xz}$ applied to the basal dislocations or by the chemical potential bias $ \delta\mu$ of the reservoirs. In the latter case the force $f_x\propto \delta \mu$ will be produced on the basal dislocation. }
	\label{fig2}
\end{figure}

{\bf Bound pairs of basal and superclimbing dislocations}.
The key element responsible for plasticity of {\it hcp} solid \he4 is the basal edge dislocation. It is characterized by Burgers vector $\vec{b}$ in the basal plane (XY-plane in Fig.~\ref{fig1}) and it can glide along this plane conservatively -- that is, without any need for extra matter injected into the bulk (see in Refs.\cite{Landau,Hirth}).
 In contrast, the superclimbing dislocation has Burgers vector $\vec{b}_c$ along the C$_6$  symmetry axis (along Z in Fig.~\ref{fig1}) and it cannot glide. However, it can climb along the basal plane with the help of extra matter supplied along its superfluid core. In Ref.\cite{sclimb} 
this process has been proposed to be responsible for the syringe effect \cite{Hallock}.  Thus, both dislocations can move along basal plane (along X in Fig.~\ref{fig1}).

A pair of basal and superclimbing dislocations interact through their elastic fields.  If their cores are parallel to each other, there are  stable equilibrium relative positions of the cores forming a straight line  (dashed lines in Fig.~\ref{fig1}) which is inclined at 45$^o$ with respect to the Burgers vectors. This line can be found  from the solution for the stress field produced by edge dislocation in isotropic medium (see in Refs.\cite{Landau,Hirth}). [The choice between two orientations of the line can be based on a simple argument that the extra half planes of both dislocations prefer not to cross each other].   
The force on the superclimbing dislocation is $f_x =b_c\sigma^{(b)}_{zz}$ (see in Refs. \cite{Landau,Hirth}) where $\sigma^{(b)}_{ij}$ is the stress tensor produced by the basal dislocation. The force on the basal dislocation is $f_x = - b \sigma^{(s)}_{zx}$, where
$\sigma^{(s)}_{ij}$ is the stress tensor due to the superclimbing dislocation. This force (its absolute value) can be found as
\beq
f_x =  \frac{|bb_c|G}{2\pi (1-\nu)} \frac{  |z(x^2-z^2)|}{(z^2+x^2)^2},
\label{fy}
\eeq 
where $G, \nu$ stand for shear modulus and the Poisson ratio, respectively; and $z,x$ define respective distances between the dislocations along Z and X axes. 

It is important to emphasize that both dislocations are confined to move along X-direction only. Thus, the distance $|z|$ along Z-axis is fixed. This allows  introducing potential energy $V(x) = - \int dx f_x$ (with its zero set at $x=0$) per unit length of the dislocations as
\beq
V(x) =  \frac{bb_c G}{2\pi (1-\nu)} \frac{zx }{z^2+x^2},
\label{Vy}
\eeq 
which features maximum and minimum at $x=\pm z$. Thus, a pair of basal and superclimbing dislocations is bound to each other with the binding energy $E=|bb_c|G/4\pi (1-\nu)$ (per unit of their length) which is independent of the distance 
between the dislocations and has a typical scale of $E\sim 10$K per atom along the core. [Distance $|z|$ between the dislocations determines the curvature of the potential energy profile as $\sim 1/z^2$]. 

An external stress $\sigma^{(ex)}_{zx}$ can break the pair apart. Indeed, such a stress will produce force $f^{(ex)} = b \sigma^{(ex)}_{zx} $  on the basal dislocation per its unit length. Thus, the potential energy of the pair will become $V^{(ex)}(x)=V(x) - f^{(ex)} x$. Formally speaking, arbitrary small $  f^{(ex)}$ can break the pair. However, there is a potential barrier for the "ionization" if $  f^{(ex)}$ is below 
some critical values $f_{cr1}$ or $f_{cr2}$ depending on the direction of the applied force. If $  f^{(ex)}$ tends to increase the distance $|x|$ between the dislocations, the threshold is    
\beq
f_{cr1} = \frac{|bb_c|G}{16\pi(1-\nu)|z|}.
\label{Vy}
\eeq 
If $  f^{(ex)}$ is applied in the opposite direction, $ f_{cr2}= 8 f_{cr1} $.
In almost ideal samples the distance $|z|$ between dislocations could be as large as few $\mu$m. Thus, the pair can be broken by a macroscopically small external stress.

{\bf Drag between basal and superclimbing dislocations}.
Applying a subcritical  force $f=b\sigma^{(ex)}_{zx}$ (along X) by external stress $\sigma^{(ex)}_{zx}$ to the basal dislocation will induce drag on the superclimbing one. This creates a chemical potential difference $\delta \mu$ between superfluid reservoirs  and the superclimbing dislocation (see in Fig.~\ref{fig2}). Accordingly, this will induce climb of the superclimbing dislocation supported by the superflow along its core. This is the syringe effect \cite{sclimb} resulting in advancing the extra plane of atoms (either up or down as sketched in Fig.~\ref{fig2}). Conversely, creating externally a difference $\delta \mu$ (by applying pressure on the reservoirs or by the Fountain effect \cite{Hallock,Moses}) will lead to injecting matter into the extra plane of atoms  which will result in the superclimb of the dislocation along X-direction (see  Fig.~\ref{fig2}). In its turn this motion will induce the force $f_x\sim \delta\mu$ on the basal dislocation causing its glide. In both cases the flow is along Y-axis while the force moving dislocations is along X-axis. This constitutes the transverse nature of the effect. 

Let's assume the dislocation network has $N_s$ paired segments of basal and superclimbing dislocations and introduce work   
\beq
W_i = - \sigma^{(i)}_{xz}b^{(i)} \xi^{(i)} 
\label{Wi}
\eeq
 (see in Ref.\cite{Hirth}) done on $i$th segment by a local stress $\sigma^{(i)}_{xz}$ moving a
basal dislocation segment of length $L_i$ by   a distance $\xi_i$ along the X-axis, where $b^{(i)}=(\vec{b}^{(i)})_x$. This stress does not affect the superclimbing dislocation directly. However, because of the dislocation pair binding, the latter will be dragged along.    
The displacement $\xi^{(i)}$ (see its direction in Fig.~\ref{fig2}) of the superclimbing dislocation along the basal plane is non-conservative and is only possible if some amount of matter $\delta N_i$ is supplied by superflow along the core (see the horizontal arrows in Fig.~\ref{fig2}). The relation between $\delta N_i$ and $\xi^{(i)}$ is of purely geometrical nature (see in Ref.\cite{Hirth}) and it depends on the sign of the Burgers vector $b^{(i)}_c= (\vec{b}^{(i)}_c)_z$ of the segment:
\beq
\delta N_i \approx \frac{L_i \xi^{(i)} b^{(i)}_c}{a^3},
\label{dN}
\eeq 
where $a\sim |b_c| \sim |b|$ is of the order of inter atomic distance.  
Here and later below numerical coefficients $\sim 1$ will be ignored.  

The tensor of plastic deformation $u^{(i)}_{xz}$ resulting from the displacement of the pair can be evaluated as
\beq
 u^{(i)}_{xz} \approx \frac{\xi^{(i)} b^{(i)}}{L_i \tilde{L}_i },
\label{upl}
\eeq 
 where $\tilde{L}_i$ is given by a typical distance between basal dislocations along Z-direction. In what follows the approximation $\tilde{L}=L_i$ and that all segments are of the same length $\tilde{L}_i=L$ will be used.  This relation simply states that displacing a basal dislocation by $L_i$ shifts the upper and lower parts of a perfect crystal between two basal dislocations by $b$ (see in Ref. \cite{Hirth}). 

It is worth mentioning that both quantities $\delta N_i$ in Eq.(\ref{dN})  and $u^{(i)}_{xz}$ in Eq.(\ref{upl}) are related to each other through the displacement $\xi^{(i)}$ of a bound pair of the basal and supetrclimbing dislocations. Thus, at least at the local level on a typical scale $\sim L$ there is a close relation between syringe effect and plastic deformation.

{\bf Syringe effect induced by plastic deformation}. 
 Let's, first, evaluate the response of $\delta N_i$ on applied uniform external stress $\sigma^{(ex)}_{xz}$. Expressing $\xi^{(i)}$ from Eq.(\ref{dN}) and substituting into Eq.(\ref{Wi}) with the replacement $\sigma^{(i)}_{xz} \to \sigma^{(ex)}_{xz}$, the work (\ref{Wi}) becomes
$W_i \approx - a^3 g_i \delta N_i \sigma^{(ex)}_{xz}$ where the notation 
\beq
g_i= \frac{b^{(i)}_c b^{(i)}}{a^2} \approx \pm 1
\label{g_i}
\eeq  
was introduced. This quantity $g_i$ varies in sign depending on the configurations of the bound dislocations shown in Fig.~\ref{fig1}.  [Given the C$_6$ symmetry of the basal plane $g_{i}$ can actually take four values: $\pm 1/2, \pm 1$]. Once particles are injected into a solid, there is a change of the compression energy which is determined by elastic moduli. This energy $W_c \approx Ka^3 (\delta N_i)^2/N_i$ where $N_i$ is a number of atoms in a volume $\sim L^3_i$ of a perfect crystal around the considered segment and $K$ stands for the compression modulus. Thus the total energy becomes
\beq
W\approx \frac{Ka^3 (\delta N_i)^2}{N_i} -  a^3 g_i  \sigma^{(ex)}_{xz}\delta N_i.
\label{Wtot}
\eeq  
Minimization of $W$ with respect to $\delta N_i$ gives the number of atoms
\beq
\delta N_i \approx g_i N_i \frac{\sigma^{(ex)}_{xz}}{K}
\label{dNsigma}
\eeq 
 injected into a volume $\sim L_i^3$ surrounding the selected pair. Within the assumption that the "conductive" network is uniform over the whole sample, Eq.(\ref{dNsigma}) applies to all $N_s$ segments.

If $g_i$ averaged over the whole sample $\langle g_i \rangle =\tilde{g}$ is non-zero, the total syringe fraction $\Delta N /N$, where $N$ stands for the total number of atoms in a sample, due to all segments would become
\beq
\frac{\Delta N}{N} = \tilde{g}\,\, \frac{\sigma^{(ex)}_{xz}}{K}.
\label{tg}
\eeq

Finite $\tilde{g}$ occurs if dislocations with a particular sign of the Burgers vectors dominate and, thus, produce global deformations.
Otherwise, $g_i$ in Eq.(\ref{g_i}) will fluctuate over a sample and should be zero if averaged over sample realizations.  However, there should be fluctuations from sample to sample leading to finite $\Delta N$ for different samples. In order to estimate the strength of the fluctuations let's introduce the mean square value of the total amount of injected atoms  $\tilde{\Delta} N =\sqrt{\langle (\sum_i \delta N_i)^2 \rangle }$, where $\langle ...\rangle $ stands for the statistical averaging over sample realizations. Using Eq.(\ref{dNsigma}),
\beq
\tilde{\Delta} N= \sqrt{\sum_{ij} \langle g_i g_j N_iN_j \rangle}\,\frac{|\sigma^{(ex)}_{xz}|}{K}. 
\label{frac0}
\eeq
As a simplest approximation,  it is reasonable to assume that the quantities $g_i$ from different segments of paired dislocations are not correlated. This implies $\langle g_i g_j\rangle =\delta_{ij}$. Then, $\sqrt{\sum_{ij} \langle g_i g_j N_iN_j \rangle} \to \sqrt{\sum_{i} \langle N^2_i \rangle}$. Considering that all segments occupy the same volume $\sim L^3$ and using $N_i\approx N/N_s$, Eq.(\ref{frac0}) becomes
\beq
\frac{\tilde{\Delta} N}{N}= \frac{1}{\sqrt{N_s}}\,\frac{|\sigma^{(ex)}_{xz}|}{K}. 
\label{frac}
\eeq
If $L_0$ is a typical sample size and $N_s \approx L_0^3/L^3$, this relation gives $\frac{\tilde{\Delta} N}{N} \approx (L/L_0)^{3/2} |\sigma^{(ex)}_{xz}|/K$.
In a sample of size, say, $L_0\sim 1$cm \cite{Hallock} with $L\sim 10\mu$m,   $\frac{\tilde{\Delta} N}{N} \approx 0.3\cdot 10^{-4} |\sigma^{(ex)}_{xz}|/K$. In smaller samples $\sim 1$mm \cite{Moses} the effect should become stronger by, at least, a factor of 30.  

{\bf The inverse syringe effect: plastic deformation induced by a superclimb}.
The reason for it also stems from the relations (\ref{upl},\ref{dN}) -- injecting a number of atoms induces a superclimb which in its turn initiates a glide of the basal dislocations bound to the superclimbing ones. 
In this case, the external bias is due to chemical potential variation applied to the SF reservoirs (see Fig.~\ref{fig2}). This leads to injecting 
of $\Delta N$ atoms into the solid. Under the assumption that the injected fraction is uniform over the solid, the relation $\delta N_i/N_i = \Delta N/N$ can be used
in Eq.(\ref{dN}). Then, $\xi_i$ can be expressed as $\xi^{(i)} \approx a b^{(i)}_c (N_i/L_i)\Delta N/N$ and then substituted into Eq.(\ref{upl}) which gives
$ u^{(i)}_{xz} \approx g_i \Delta N/N$ where the relation $N_i \approx L_i^2 \tilde{L}_i /a^3$ has been used. Thus, if the sample average $\tilde{g}$ of $g_i$ is finite, the global shear of the sample becomes
\beq
 u_{xz} \approx \tilde{g} \,\, \frac{\Delta N}{N},
\label{uav}
\eeq
which is the inverse version of the relation (\ref{tg}).
If, however, $\tilde{g}=0$, the mean square fluctuation $\tilde{\Delta} u_{xz}$ of the shear deformation becomes
 \beq
\tilde{\Delta} u_{xz}\approx \frac{1}{\sqrt{N_s}} \frac{|\Delta N|}{N} \to \left(\frac{L}{L_0}\right)^{3/2} \frac{|\Delta N|}{N}.
\label{globalU}
\eeq
Choosing typical $|\Delta N|/N \sim 0.01-0.1$ observed in Ref.\cite{Hallock} and the same values of $L_0, L$ as above, the 
magnitude of the strain fluctuations (from sample to sample) becomes $\sim 10^{-6} - 10^{-5}$. These values are well within the range detectable in the setup \cite{Beamish_strain} where strains as low as $\sim 10^{-9}$ have been  observed.   

{\bf Polycrystalline \he4 }. 
In polycrystaline samples with random orientations of grains the described transverse supershear can only be observed with respect to fluctuations of $\Delta N/N$ and shear strain tensor  $u_{\alpha\beta}$ (where the indices $\alpha, \beta$ refer to the $X,Y,Z$ directions)  in the direct and inverse versions, respectively. Averaging Burgers vectors $\vec{b}$ and $\vec{b}_c$ of the intra-grain dislocations over grains in a given sample may produce non-zero tensor  $g_{\alpha \beta}= \langle (b_c)_\alpha b_\beta \rangle/a^2$, with $ g_{\alpha \alpha}=0$ due to  $\vec{b}_c \vec{b}=0$ (with the summation performed over repeated indices).  [This tensor becomes zero after averaging over sample realizations].  
Then, e.g.  for the inverse effect the shear strain (in a given sample) is determined as $u_{\alpha \beta} \approx \kappa (g_{\alpha \beta} + g_{\beta \alpha })\Delta N/N$ where $\kappa$   is a numerical coefficient determined by a mismatch between gliding planes of neighboring grains. In general, it should be $\kappa <<1$. Thus, the fluctuation of the plastic strain should be reduced by the factor $\kappa$, if compared with the relation (\ref{globalU}).  
 The direction of the flow
producing the syringe effect is set along the direction $\pm \varepsilon_{\alpha \beta \gamma} g_{\beta \gamma}$, where $\varepsilon_{\alpha \beta \gamma}$ stands for the Levi-Civita symbol.

{\bf Discussion}. 
The discussed effects should be realized in the geometry sketched in Fig.~\ref{fig2}. Namely, the shear stress
must be applied perpendicular to the direction of the superflow between the reservoirs. Furthermore, the resulting force on the dislocations should be along the basal plane. Thus, the optimal condition is to have a single crystal with known orientation of the C-axis. 
[As the symmetry analysis conducted above shows, the effects should also exist in polycrystalline samples -- albeit in its reduced form].

It is worth mentioning that the above estimate for $N_s \sim (L_0/L)^3$ in Eqs.(\ref{globalU},\ref{frac}) is actually too conservative. Since the orientation of Burgers vector does not change as dislocation line meanders through the solid from one reservoir to another, the value of $g_i$, Eq.(\ref{g_i}), may persist over the whole length $\sim L_0$. This, then, will give the number of segments scaled as $N_s \sim (L_0/L)^2$ and will increase the above estimates for the fluctuations of the responses (\ref{frac},\ref{globalU}) by a factor of 30.

Introducing basal and superclimbing dislocations with prevalence of the corresponding Burgers vectors of one sign will enhance the effects as determined by the tensor $\tilde{g}$ in Eqs.(\ref{tg},\ref{uav}). This can be achieved by growing crystal in a geometry introducing basal mismatch dislocations of a particular sign inducing global rotation of the basal plane determined by  the angle $ \sim b/d_b$, where $d_b$ stands for the inter-dislocation distance.  Similarly, injecting atoms from only one side of a sample will introduce superclimbing dislocations of definite sign characterized by the mean separation $d_{sc}$. Then, keeping in mind the pairing between these dislocations, one of the four configurations shown in Fig.~\ref{fig1} will dominate, and, thus,    $\tilde{g}$ can be determined as $| \tilde{g}| \sim [{\rm min}( b/d_b,  b_c/d_{sc})]^2$.   

It should be mentioned that bound complexes of more than two dislocations can be formed. This, while making the analysis more involved, does not change the results.

 {\bf Acknowledgment}. I thank Nikolay Prokof'ev and Boris Svistunov for useful discussions.
This work was supported by the National Science Foundation under the grant DMR-1720251.

\end{document}